\documentclass[12pt,a4paper]{article}
\usepackage{amsmath,amssymb,amsthm,amsfonts}
\pdfoutput=1
\usepackage{geometry}
\usepackage{cite}
\usepackage{graphicx}
\usepackage{epsfig}
\usepackage{cases}
\usepackage{caption}
\usepackage{subcaption}
\usepackage[pagebackref=false,pageanchor=false]{hyperref}
\usepackage[affil-it]{authblk}
\hypersetup{colorlinks=true,linkcolor=blue,filecolor=magenta,urlcolor=cyan,}
\usepackage[title]{appendix}
\numberwithin{equation}{section}
\def\a{\alpha}
\def\b{\beta}
\def\g{\gamma}
\def\d{\delta}

\def\z{\zeta}

\def\k{\kappa}
\def\l{\lambda}
\def\m{\mu}
\def\n{\nu}

\def\r{\rho}

\def\s{\sigma}

\def\Gg{{\mathcal G}}

\def\P{\Pi}

\def\G{\Gamma}

\newcommand{\nn}{\nonumber}
\def\be{\begin{equation}}
\def\ee{\end{equation}}
\def\bea{\begin{eqnarray}}
\def\eea{\end{eqnarray}}
\def\bg{\begin{align}}
\def\eg{\end{align}}
\def\bs{\begin{split}}
\def\es{\end{split}}

\def\pa{\partial}

\def\td{\tilde}

\def\hlf{\frac{1}{2}}

\def\ie{{\it i.e. }}

\def\eg{{\it e.g. }}

\begin{document}
\begin{titlepage}
{\title{{A Holographic Study of the $a$-theorem  and RG Flow in General Quadratic Curvature Gravity}}}
\vspace{.5cm}
\author{Ahmad Ghodsi\thanks{a-ghodsi@ferdowsi.um.ac.ir}}
\author{Malihe Siahvoshan\thanks{malihe.siahvoshan@mail.um.ac.ir}}
\vspace{.5cm}
\affil{ Department of Physics, Faculty of Science,\\ 
	Ferdowsi University of Mashhad,  Mashhad, Iran}
\renewcommand\Authands{ and }
\maketitle
\thispagestyle{empty}
\vspace{-12cm}
\begin{flushright}
%{\small
%{\tt arXiv:yymm.nnnn} \\
%}
\end{flushright}
\vspace{10cm}
%%%%%%%%%%%%%%%%%%%%%%%%%%%%%%%%%%%%%%%%%%%%%%%%%%%%%%%%%%%
\begin{abstract} 
We use the holographic language to show the existence of the $a$-theorem for even dimensional CFTs, dual to the AdS space in general quadratic curvature gravity. We 
find the Wess-Zumino action which is originated from the spontaneous breaking of the conformal symmetry in  $d\leq 8$, by using a radial cut-off near the AdS boundary. We also study the RG flow and (average) null energy condition in the space of the couplings of theory. In a simple toy model, we find the regions where this holographic RG flow has a monotonic decreasing behavior.
\end{abstract}

\end{titlepage}
%%%%%%%%%%%%%%%%%%%%%%%%%%%%%%%%%%%%%%%%%%%%%%%%%%%%%%%%%%%
%\setcounter{footnote}{0}
%\addtocontents{toc}{\protect\setcounter{tocdepth}{4}}
%\setcounter{secnumdepth}{4}
%\tableofcontents
%%%%%%%%%%%%%%%%%%%%%%%%%%%%%%%%%%%%%%%%%%%%%%%%%%%%%%%%%%%%
%%%%%%%%%%%%%%%%%%%%%%%%%%%%%%%%%%%%%%%%%%%%%%%%%%%%%
\section{Introduction}
In the context of two-dimensional unitary conformal field theories, the Zomolodchikov's $c$-theorem \cite{Zamolodchikov:1986gt} states that the central charge monotonically decreases along the Renormalization Group (RG) flow. 
We can expect this from the Wilsonian approach in quantum field theory, in which, by integrating out the high energy modes, the number of degrees of freedom decreases.
Komargodski and Schwimmer have proved a generalization of this theorem in \cite{Komargodski:2011vj, Komargodski:2011xv}, after the conjecture of Cardy \cite{Cardy:1988cwa}. They prove an $a$-theorem for four-dimensional unitary conformal field theories and show that for any RG flow between a UV and an IR fixed point $a_{UV}\geq a_{IR}$. Here $a_{UV}$ and $a_{IR}$ are the coefficients of the four dimensional UV/IR conformal anomaly, which can be computed from the non-vanishing value of the trace of the energy-momentum tensor 
\be
\left\langle {T^{\mu}}_{\mu}\right\rangle= c\, C_{\mu\nu\a\b}C^{\mu\nu\a\b}-a E_4\,,
\ee
where $C$ is the Weyl tensor and $E_4$ is the Euler density (Gauss-Bonnet terms in four dimension).
In the proof of \cite{Komargodski:2011vj}, there is a Nambu-Goldstone boson $\sigma$, corresponding to the spontaneously broken conformal symmetry and an effective action $W[\sigma]$, which is emerging by integrating out the degrees of freedom along the RG flows driven by adding the relevant operators.

In \cite{Sinha:2012tc} the same idea is investigated in the context of the AdS/CFT. They holographically construct the $W[\sigma]$ and follow its changes along the RG flow. To find the effective action, they start from the gravity side by considering a bulk action together with the Gibbons-Hawking (GH) terms and counter-terms. The other ingredients are the AdS metric in the Poincare coordinate (flat boundary space) together with a radial cut-off near the AdS boundary. The later plays the role of the RG scale. By promoting $z$, the radial coordinate of the cut-off surface, to $\sigma$ as a  (spurion) field, \ie $z=e^{\s}$, and by computing the bulk and boundary actions, after a derivative expansion one will find a Wess-Zumino (WZ) action for the spurion field in the even dimensions. This effective action directly is related to the conformal anomaly in even dimensions as discussed in \cite{Sinha:2012tc}. 

In this holographic approach, depending on which AdS throat we are dealing with, the coefficient of the effective dilaton action is equal to the value of $a_{UV}$ or $a_{IR}$. These AdS solutions correspond to the UV/IR fixed points of the RG flow. When one considers the contributions of both throats, the overall coefficient would be $a_{UV}-a_{IR}$, which it has been proved  in the quantum field theory side \cite{Hoyos:2012xc}. The study of the dilaton WZ effective action in even $d$ dimensions and up to and including the 8-derivative terms has been performed in \cite{Elvang:2012yc}.

The generalization of the $c$-theorem to higher dimensions mostly has been done by using the gauge/gravity correspondence. For the early works including the holographic renormalization and RG flows through the relevant deformations of the Lagrangian see \cite{Akhmedov:1998vf, Balasubramanian:1999jd, deBoer:1999xf, Bianchi:2001de ,Bianchi:2001kw, Freedman:1999gp, Girardello:1999bd, Bigazzi:2001ta, Martelli:2001tu, Berg:2001ty, D'Hoker:2002aw, Freedman:2003ax}.

Another direction for the generalization of the holographic $c$-theorem is the extension of the bulk Lagrangian to the higher curvature terms. The first attempts have been done in \cite{Myers:2010xs, MyersSinha:2010tj} for quasi-topological gravities and in \cite{LiuSabraZhao:2010xc} for Lovelock and $f(R)$ theories of gravity. Unlike the four dimensional holographic CFT dual to the AdS solution in the Einstein gravity, in quasi-topological theories $a\neq c$. In these theories, it is possible to show that for a general RG flow there is a monotonically decreasing function $a(r)$, assuming that the matter sector obeys the null energy condition. This function at the fixed points reproduces correct values for $a_{UV}$ and $a_{IR}$. With the same conditions, one cannot find a similar function for $c(r)$, \cite{Myers:2010xs}. 
In this direction, an $a$-function for four dimensional general curvature square gravity  has been found in \cite{BhattacharjeeSinha:2015qaa}. The non-increasing behavior of this function is proved by using the Raychaudhuri equation. The non-increasing RG flow is restricted to a certain class of curvature square theories.

In \cite{Banerjee:2015uaa} an a-function is introduced by using the Jacobson-Myers (JM) entropy functional. The non-increasing behavior of this function follows from the fact that the JM entropy functional satisfies the linearized second law of the causal horizon thermodynamics. This study includes the general curvature squared gravity and $f(R)$ gravity. It also shows that in the absence of the null energy condition for certain theories which a scalar field is coupled to the gravity in AdS space, the second law would be enough condition for the monotonicity.

Further study including the Ricci polynomials in the bulk Lagrangian is presented in \cite{Li:2017txk}. They show the existence of an $a$-theorem for the Ricci cubic theory by restricting the couplings of the theory. These constraints are inconsistent with the ghost-free condition of the theory, but at the level of the Riemann cubic theories, the constraints for non-increasing RG flow coincide with the ghost-free conditions.
For further studies of the cubic gravities see also \cite{Bueno:2018xqc} and \cite{Li:2019auk}.\footnote{For recent studies in holographic RG flow also see \cite{Ghosh:2017big, Alkac:2018whk, Park:2018hfd}.}

In this paper, we are going to study the holographic $a$-theorem for general quadratic curvature (GQC) gravity following the reference \cite{Sinha:2012tc}. In section 2, we use the perturbative method for maximally symmetric solutions to find an effective action for the GH terms. We also use the known algorithm for finding the counter-terms. These terms are sufficient to cancel the divergences of quadratic curvature gravity with dimension less than ten. By finding these terms, we are able to compute the related WZ actions in even dimensions. We also read the corresponding coefficients to find the value of the $a$-charge.

In section 3 we study the holographic RG flow between the UV and IR fixed points. We use a kink solution, which we suppose it to satisfy the equations of motion in the presence of the matter field. This solution reduces to the AdS solution at both UV/IR limits. We use it to study the behavior (monotonicity) of the RG flow.
We suppose a proper ansatz for the holographic RG flow with general coefficients and find the possible regions in the space of couplings where the value of this RG flow monotonically decreases. We also check the regions where the (average) null energy condition holds. In the last section, we summarize our computations and discuss the results.

%%%%%%%%%%%%%%%%%%%%%%%%%%%%%%%%%%%%%%%%%%%%%%%%%%
\section{Dilaton action in GQC gravity}
\label{GeneralQuadraticAction}
In the context of the gauge/gravity correspondence, the reference \cite{Sinha:2012tc} finds the effective dilaton actions corresponding to the spontaneously broken conformal symmetry in even dimensions. In this section, we are going to generalized the idea in \cite{Sinha:2012tc} to find the dilaton actions in GQC gravity. 
It holographically guarantees the existence of an $a$-theorem for the dual gauge theories.
To construct the effective dilaton action, we  begin from the following total action
\be\label{R2TotalAction}
S_{tot}= S_{bulk}+S_{GH}+S_{ct}\,.
\ee
This action is constructed from the $d+1$ dimensional bulk action plus the boundary parts, including the GH terms and counter-terms. 
In the following subsections, we are using the standard known algorithms, to compute the corresponding GH terms and the counter-terms of the GQC gravity in the Euclidean signature. The bulk action of the theory is
\be
\label{R2action}
S_{bulk}=-\frac{1}{2\k^2}\int_{\mathcal{M}}d^{d+1}x\sqrt{g}\Big(\frac{d(d-1)}{L^2} + R +\a_1 R^2 + \a_2 R_{\m\n}^2 + \a_3 R_{\m\n\r\s}^2 \Big).
\ee
The first two terms in this Lagrangian are the familiar cosmological and Einstein-Hilbert terms. In a special case which $\a_2=-4\a_1=-4\a_3$, the quadratic part of the Lagrangian is the well-known Gauss-Bonnet (GB) terms.  
We can redefine the curvature squared coefficients in a way that the action is written as a combination of the GB terms plus the Ricci squared terms \ie
\be
\label{R2actionWith a3}
S_{bulk}=-\frac{1}{2\k^2}\!\int_{\mathcal{M}}\!\!d^{d+1}x\sqrt{g}\Big(\frac{d(d-1)}{L^2}+ R +\mathit{a}_1 R^2 + \mathit{a}_2 R_{\m\n}^2 + \mathit{a}_3 (R^2 - 4 R_{\m\n}^2+R_{\m\n\r\s}^2)\Big).
\ee
This action admits an AdS solution, $R_{\m\n}=-\frac{d}{\td{L}^2}\,g_{\m\n}$, where $\td{L}$ is the radius of AdS space-time and is related to the cosmological parameter $L$ by the following equation
\be
\label{LT}
\frac{1}{\tilde{L}^2} = \frac{1}{{L}^4} \Big( {L}^2 + \frac{d-3}{d-1} \big( d(d+1) \mathit{a}_1 + d\mathit{a}_2 + (d-1) (d-2)\mathit{a}_3\big) \Big).
\ee

\subsection{Gibbons-Hawking surface terms}
A challenge in the study of the higher order curvature theories of gravity is the computation of the GH terms. These surface terms are necessary to have a well-defined variational principle. In the Einstein-Hilbert (EH) gravity and in the Lovelock theories of gravity, these terms are well-known \cite{Myers:1987yn}. For the EH gravity, the GH term is given by
\be\label{GHEH}
S_{GH}^{EH}=-\frac{1}{\k^2}\int_{\pa\mathcal{M}}\!\! d^dx \, \sqrt{h} \, K,
\ee
where $h$ is the determinant of the induced boundary metric and $K$ is the trace of the extrinsic curvature of the boundary surface, $\partial\mathcal{M}$. The extrinsic curvature is defined by $K_{\mu\nu}=2\nabla_{(\m} n_{\n)}$ and $n_\m$ is the space-like unit vector normal to the boundary.
Moreover, there is a generalized GH action for the GB gravity \cite{Myers:1987yn}
\be
\label{gb gh}
S_{{GH}}^{GB}=\frac{2}{\k^2} \mathit{a}_3\int_{\pa\mathcal{M}}\!\!d^dx \, \sqrt{h} \,\Bigl( 2 \Gg_{ab} {K}^{ab} +\tfrac{1}{3}(K^3 - 3 K K_{ab} K^{ab} + 2 K_{ab} K^{bc} K_{c}{}^{a} ) \Bigr),
\ee
where $\Gg_{ab}$ is the Einstein tensor constructed out of the induced boundary metric.

In general, to find the related GH terms for the remaining Ricci curvature terms, $\mathcal{L}^{R^2}=\mathit{a}_1 R^2 + \mathit{a}_2 R_{\m\n}^2$, the usual method does not work, \ie by variation of the Lagrangian with respect to the metric one cannot find the suitable terms to have a well-defined variational principle. 
However, according to the perturbative method in \cite{Cremonini:2009ih}, for a maximally symmetric solution, we can find an effective GH term.

In variation of the EH action, the corresponding surface terms which are coming from $\sqrt{g}\,g^{\m \n}\d R_{\m\n}$ part, are eliminated by the GH term in (\ref{GHEH}). 
We can use this fact to compute the effective GH terms for $\mathcal{L}^{Ric}=\mathcal{L}^{EH}+\mathcal{L}^{R^2}$, as far as we are working  in a  maximally symmetric background. By variation with respect to the Ricci tensor we find
\be
\begin{split}
\frac{\d \mathcal{L}^{Ric}}{\d R_{\m\n}} =   (1+2\mathit{a}_1 R) \mathit{g}^{\m\n}+ 2 \mathit{a}_2 R^{\m\n}\,.
\end{split}
\ee
We are just interested  to keep  up to the first power of the couplings of the theory in the effective Lagrangian, therefore we can substitute the zero order solution \ie $ R_{\m\n} = -\frac{d}{L^2}\, g_{\m\n}$ into the above equation. 
Finally, the related GH term is the GH term of the EH gravity with an effective coefficient, \ie
\be
\label{r2 gh}
S^{Ric}_{GH}\equiv\l_{eff} \,S_{GH}^{EH}= -\frac{1}{\k^2}\big(1 - \frac{2\mathit{a}_1}{L^2}\, d(d+1) - \frac{2\mathit{a}_2}{L^2}\, d \big)\,\int_{\pa\mathcal{M}}\!\! d^d x\,\sqrt{h}\,K .
\ee
Consequently, the total generalized GH surface terms for the GQC gravity will be the sum of the GH terms in \eqref{gb gh} and  \eqref{r2 gh}.
%%%%%%%%%%%%%%%%%%%%%%%%%%%%%%%%%%%%%%%%%%%%%%%%%%%
\subsection{Counter-terms}
To make a finite total on-shell action, we need some additional boundary counter-terms.
The counter-terms of the GB part of the action has already been calculated in \cite{Liu:2008zf} (see also \cite{Yale:2011dq})
\begin{align}\label{LctGB}
\mathcal{L}_{ct}^{GB}= &-\mathit{a}_3 \bigg\{\frac{(d-2)(d-3)}{L^2} \bigl(\frac{d-1}{6L} +  \frac{3\,L}{4\,(d-2)}\, \mathcal{R} \nn \\ 
&+\frac{7\,L^3}{4\,(d-2)^2\,(d-4)}(\mathcal{R}_{ab}\mathcal{R}^{ab}-\frac{d}{4 (d-1)}\mathcal{R}^2) \bigr) \nn \\
& - \frac{L}{d-4} \big(\mathcal{R}^2 -4\,\mathcal{R}_{ab}\mathcal{R}^{ab}+\mathcal{R}_{abcd}\mathcal{R}^{abcd}\big) +\cdots\bigg\}\,,
\end{align}
where all the curvature tensors are constructed from the induced metric on the boundary.

In the following we are going to use the suggested algorithm in \cite{Kraus:1999di} to find all the proper counter-terms of the Ricci square terms of the bulk action. 
According to \cite{Brown:1992br}, the energy-momentum tensor related to the EH action is defined by
\be\label{EMD}
\P^{ij} \equiv \frac{2}{\sqrt{h}}\frac{\d S^{EH}}{\d h_{ij}} = K^{ij} - K h^{ij}\,,
\ee
this contains a divergent part on the boundary. A basic motivation for adding the counter-terms is the elimination of this divergent part, in order to have a finite energy-momentum tensor. 
In this regard, we demand that the constructed energy-momentum tensor from the counter-term action
\be\label{pict}
\td{\P}^{ij} = \frac{2}{\sqrt{h}}\frac{\d S_{ct}}{\d h_{ij}} = \frac{2}{\sqrt{h}}\frac{\d}{\d h_{ij}} \int_{\pa\mathcal{M}}\!\! d^d x\,\sqrt{h}\,\mathcal{L}_{ct} \,,
\ee
cancels out the divergent part of the energy-momentum tensor of (\ref{EMD}).
On the other hand, the effective GH term of (\ref{r2 gh}) suggests that for the Ricci squared part of the action the extrinsic curvature must be re-scaled by a factor of $\l_{eff}$. Therefore the new energy-momentum tensor must be equal to the (\ref{EMD}) expression with an extra factor of $\l_{eff}$. 
By this assumption, the Gauss-Codazzi (GC) equation would change to
\be
\label{GCE}
2\,G_{\m\n} n^\m n^\n = \frac{\P^2}{\l_{eff}^2(d-1)} - \frac{\P^{ab}\P_{ab}}{\l_{eff}^2} - \mathcal{R}\,,
\ee
where $G_{\m\n}$ is the Einstein tensor constructed from the bulk metric.
The main difference between the work here and the one in \cite{Kraus:1999di} is the existence of the  $\l_{eff}$ coefficients in the GC equation \eqref{GCE}. We must also remember that for the general quadratic action of (\ref{R2actionWith a3}), the radius of the AdS solution, $\td{L}$, is given by \eqref{LT}.
By considering the AdS solution for the bulk Lagrangian, the left hand side of \eqref{GCE} is equal to
\be
2\,G_{\m\n} n^\m n^\n =  \frac{d(d-1)}{\td{L}^2}\,.
\ee
Now the GC equation \eqref{GCE} can be rewritten for the divergent part of the energy-momentum tensor.  The relevant GC equation would be
\be
\label{GCE2}
\frac{d(d-1)}{\td{L}^2} \l_{eff}^2 + \mathcal{R}\, \l_{eff}^2 + \td{\P}^{ab}\td{\P}_{ab} - \frac{ \td{\P}^2}{(d-1)}  = 0\,.
\ee
To solve this equation perturbatively, $\td{\P}_{ab}$ can be expanded in terms of the AdS radius as
\be
\td{\P}_{ab}=\td{\P}^{(0)}_{ab}+\td{\P}^{(1)}_{ab}+\td{\P}^{(2)}_{ab}+\cdots \,.
\ee
In the calculation of the zero order of $\td{\P}$, the curvature $\mathcal{R}$ does not contribute, therefore $\td{\P}^{(0)}_{ab}$ should be proportional to the boundary metric, \ie $\td{\P}^{(0)}_{ab}= \a h_{ab}$. Inserting this  into the  equation (\ref{GCE2}), one finds
\be
\a = - \frac{(d-1)\,\l_{eff}}{\td{L}} \,.
\ee
The Weyl symmetry of the theory gives a freedom to write the following relation between the trace of the energy-momentum tensor and the corresponding counter-term in all orders \cite{Kraus:1999di}
\be
\label{weyl}
(d-2n) \, \td{\mathcal{L}}^{(n)} = \td{\P}^{(n)} \,.
\ee
Altogether, for the zeroth order of the energy-momentum tensor we have
\begin{subequations}
\begin{align}
\td{\P}^{(0)} &= - \frac{d(d-1)\,\l_{eff}}{\td{L}}\,,\\
\text{(\ref{weyl})} \Rightarrow \td{\mathcal{L}}^{(0)}&= - \frac{(d-1)\,\l_{eff}}{\td{L}}\,,\\
\text{(\ref{pict})} \Rightarrow \td{\P}^{(0)}_{ab} &= -\frac{(d-1)\,\l_{eff}}{\td{L}}~h_{ab} \,.~~~~~~~~~~~~~~~~~~~~~~~
\end{align}
\end{subequations}
We can write the final result in terms of the original Lagrangian coefficients as follow
\be\label{Lct0}
\td{\mathcal{L}}^{(0)}= \frac{d-1}{L} - \mathit{a}_1 \frac{d(d+1)(3d-1)}{2 L^3} - \mathit{a}_2 \frac{d(3d-1)}{2 L^3}\,.
\ee
Although it is the beauty of this method that it seems the signature of the higher order corrections becomes evident in an overall factor $\l_{eff}$ and the new AdS radius $\td{L}$, but this argument is correct as far as the higher curvature terms in the bulk action are made of the Ricci tensor.
Having the zero order, we can repeat the above steps of the algorithm for the next orders of the energy-momentum tensor. 

Up to the cubic curvature terms we find the following counter-terms
\begin{subequations}
\begin{align}\label{Lct121}
\td{\mathcal{L}}^{(1)} &= - \l_{eff} ~ \frac{\td{L}}{2 \,(d-2)}  \, \mathcal{R}\,,\\
\label{Lct122}
\td{\mathcal{L}}^{(2)} &= - \l_{eff} ~ \frac{\td{L}^3}{2\, (d-2)^2\,(d-4)}\Bigl(\mathcal{R}_{ab} \mathcal{R}^{ab} - \frac{d}{4 (d-1)}\mathcal{R}^2\Bigr)\,,\\
\label{Lct123} 
\td{\mathcal{L}}^{(3)} &=  \l_{eff} ~  \frac{\td{L}^5}{(d-2)^3(d-4)(d-6)} \Big(\, \frac{3d+2}{4(d-1)} \mathcal {R}\mathcal{R}_{ab} \mathcal{R}^{ab} -  \frac {d(d+2)} {16 (d-1)^2}\mathcal {R}^3  \\
&~~~-2 \mathcal {R}^{ab} \mathcal {R}^{cd} \mathcal {R}_{acbd} + \frac{d-2}{2(d-1)} \mathcal{R}^{ab} \mathcal{D}_{a} \mathcal{D}_{b} \mathcal {R}
+ \frac{1}{2 (d-1)} \mathcal{R} \mathcal{D}_{a}\mathcal{D}^{a} \mathcal{R} - \mathcal{R}^{ab} \mathcal{D}_{c}\mathcal{D}^{c} \mathcal{R}_{ab} \Big)\,.\nn
\end{align}
\end{subequations}
The first and the second lines, \eqref{Lct121} and \eqref{Lct122},  exactly reproduce the results of \cite{Cremonini:2009ih} when the value of $\l_{eff}$ is inserted from equation \eqref{r2 gh}.
Finally, the counter-terms which are sufficient to cancel the divergences of the GQC gravity in AdS space for $d<10$, are the sum of (\ref{LctGB}), (\ref{Lct0}) and (\ref{Lct121})-(\ref{Lct123}) Lagrangians
\be\label{Sct1}
\begin{split}
S_{ct} &= \frac{1}{2\k^2}\int_{\pa\mathcal{M}}d^d x\,\sqrt{-h}\,(\mathcal{L}_{ct}^{GB} + \td{\mathcal{L}}^{(0)} + \td{\mathcal{L}}^{(1)} + \td{\mathcal{L}}^{(2)}+ \td{\mathcal{L}}^{(3)})\,.\\
\end{split}
\ee
%%%%%%%%%%%%%%%%%%%%%%%%%%%%%%%%%%%%%%%%%%%%%%%%%%%%%%%%%%%%%
\subsection{a-theorem in GQC gravity}
\label{R2a-Theorem}
By finding the GH terms and counter-terms we are able to follow the next step in \cite{Sinha:2012tc} to prove the $a$-theorem for GQC gravity holographically.  We consider the Euclidean AdS metric in the Poincare coordinate as follow
\be
\label{EuclideanAdS}
ds^2 = \td{L}^2 \, \frac{dz^2 + \d_{ab} \,dx^a dx^b}{z^2}\,,\qquad a, b=1, \cdots, d\,.
\ee
The curvature tensors constructed from this metric satisfy  the following relations
\be
R_{\m\n\a\b} = -\frac{1}{\td{L}^2} \, (g_{\m\a}g_{\n\b}-g_{\m\b}g_{\n\a})\,,~~~~~~R_{\m\n} = -\frac{d}{\td{L}^2} \,g_{\m\n}\,,~~~~~~R = -\frac{d\,(d+1)}{\td{L}^2}\,.
\ee
Moreover, we introduce a radial cut-off  as a scalar function of the boundary variables through $z = e^{\s(x^a)}$ which plays the role of the RG scale.
At the position of this cut-off surface, the induced metric is given by
\be
\label{BoundaryMetric}
h_{ab} = \td{L}^2 \,(\pa_a \s \pa_b \s + e^{2\s}\, \d_{ab} )\,.
\ee
We now substitute the above metric into the total action of \eqref{R2TotalAction}  and expand the result in terms of the derivatives of the scalar field. We expect that similar to the GB case in \cite{Sinha:2012tc}, if there exists an $a$-theorem for the GQC theory, then the final scalar action is a WZ action with an overall coefficient proportional to the conformal anomaly in even dimensions.

 To accomplish this, we need to find the intrinsic and extrinsic curvatures constructed from the boundary metric of \eqref{BoundaryMetric}. All the computations related to this subsection are presented in Appendix A.
The results of our calculations for various even dimensions are listed below:

\subsubsection{\texorpdfstring{$\textbf{d=2}$}{TEXT}}
In two dimensions, the GB part of the action \eqref{R2actionWith a3} together with its corresponding GH terms and counter-terms do not contribute. The total action is
\be
S_2 = - \frac{(\tilde{L}^2 - 12 \mathit{a}_1 - 4 \mathit{a}_2 )}{2\k^2 \tilde{L}} \int d^2x \Bigl( (\pa \s)^2 + 2 \Box \s \Bigr)\,,
\ee
where $\Box=\partial_a\partial^a$. 
Dropping the total derivative term, $S_2$ is a two dimensional WZ action
\be
S_2 = \frac{a^*_2}{\pi}  \int d^2x ~ \mathcal{L}^{(2)}_{WZ}\,,
\ee
where
\be
\mathcal{L}^{(2)}_{WZ}= - \hlf (\pa \s)^2\,, \qquad  \qquad
a^*_2 = (\tilde{L}^2 - 12 \mathit{a}_1 - 4 \mathit{a}_2 )\, \frac{\pi}{\k^2\tilde{L}}\,.
\ee
\subsubsection{\texorpdfstring{$\textbf{d=4}$}{TEXT}}
In this dimension all terms, including the related GB term, must be considered. 
After the derivative expansion, we drop the surface terms that are generated from  by part integration at the fourth order, therefore, the total action reduces to
\be
\begin{split}
S_4 &= \frac{\tilde{L}}{2\k^2}  \bigl( \tilde{L}^2 -4 (10 \mathit{a}_1 + 2 \mathit{a}_2 + 3 \mathit{a}_3) \bigr) \!\!\int \!\! d^4x \Big(\pa (e^{-2\s}\pa \s) 
+ \frac{1}{4} \big( 2\Box \s(\pa \s)^2  - (\pa \s)^4 \big) \Big)\,.
\end{split}
\ee 
The second order derivative term is a total derivative too, so the final result is
\be
S_4 = \frac{a^*_4}{\pi^2}  \int d^4x ~\mathcal{L}^{(4)}_{WZ}\,,
\ee
where 
\be
\mathcal{L}^{(4)}_{WZ} = \frac{1}{4} \, \Box \s (\pa \s)^2 - \frac{1}{8} \,(\pa \s)^4\,,\qquad a^*_4 = \frac{\pi^2}{\k^2}\tilde{L} \,\Bigl( \tilde{L}^2 -4  \,(10 \mathit{a}_1 + 2 \mathit{a}_2 + 3 \mathit{a}_3) \Bigr) \,.
\ee

\subsubsection{\texorpdfstring{$\textbf{d=6}$}{TEXT}}
Up to the 6th order of the derivatives, in six dimension the action is given by
\begin{align}
S_6 = & \frac{\tilde{L}^3}{\k^2} \,\bigl( \tilde{L}^2 - 4(21 \mathit{a}_1 + 3 \mathit{a}_2 + 10 \mathit{a}_3) \bigr) \int d^6x \Bigl( \frac{1}{4}  \pa  (e^{-4\s}\pa \s)+ \frac{1}{8}  \pa_a  (e^{-2\s} \pa_b \s \pa^b \s \pa^a \s)   \cr 
& - \frac{1}{32} \big( 2(\pa \s)^6 - 3 (\pa \s)^4 \pa^2 \s
 + 2(\Box\s)^2(\pa \s)^2  - 2 (\pa \s)^2 (\pa\pa \s)^2 \big) \Bigr)\,.
\end{align}
After eliminating all the remaining surface terms in the second and fourth order of the derivatives, the final result would be a six derivative action
\be
S_6 = \frac{a^*_6}{\pi^3}  \int d^6x ~ \mathcal{L}^{(6)}_{WZ}\,,
\ee
where
\be
\begin{split}
\mathcal{L}^{(6)}_{WZ} &= -\frac{1}{8} \,(\pa \s)^6 + \frac{3}{16}\, \Box \s(\pa \s)^4  - \frac{1}{8}\,(\Box\s)^2 (\pa \s)^2 + \frac{1}{8}\, (\pa \s)^2 (\pa\pa \s)^2\,,\\
a^*_6 &= \frac{\pi^3\tilde{L}^3}{2\k^2} \,\Bigl( \tilde{L}^2 - 4(21 \mathit{a}_1 + 3 \mathit{a}_2 + 10 \mathit{a}_3) \Bigr) \,.
\end{split}
\ee

\subsubsection{\texorpdfstring{$\textbf{d=8}$}{TEXT}}
The calculations at this dimension are not very straightforward like the former cases. This is because the dilaton action consists of a WZ part plus some non-vanishing Weyl invariant terms in this dimension \cite{Elvang:2012yc}. To simplify the computations we drop the GB terms to see just the effects of the Ricci squared terms of the action \eqref{R2actionWith a3} \footnote{To evaluate the 8-dimensional total action we would need the proper GB counter-terms, which in principle its calculation is possible due to earlier proposed methods, \eg \cite{Liu:2008zf}}.
There are nine Weyl invariant terms in 8 dimensions, in other words the dilaton action can be written as \cite{Elvang:2012yc}
\begin{align}
\label{CompeleteS8}
S_8 =& \int d^8x ~\sqrt{-h} ~\Bigl(  \g_1 \mathcal{R}^4 + \g_2 \mathcal{R}^2 \mathcal{R}_{ab}^2 + \g_3 \mathcal{R} \mathcal{R}_{ab}^3 +\g_4 (\mathcal{R}_{ab}^2)^2 +\g_5 \mathcal{R}_{ab}^4 + \g_6 (\square \mathcal{R})^2 \nn \\
&+ \g_7 (\square \mathcal{R}_{ab})^2 + \g_8 \mathcal{R}(\mathcal{D}_a \mathcal{R})^2 + \g_9 \mathcal{R}_{ab}^2 \square \mathcal{R} \Bigr) + \frac{a^*_8}{\pi^4}  \int d^8x ~ \mathcal{L}^{(8)}_{WZ}\,,
\end{align}
where $\mathcal{L}^{(8)}_{WZ}$ is the 8 dimensional Wess-Zummino action 
\begin{align}
\mathcal{L}^{(8)}_{WZ}= &-\frac{1}{64}\!\int \!d^8x  \Bigl( 3 \Box^2 \s (\pa \s)^2 + 6 (\Box \s)^3 (\pa \s)^2 + 36 (\Box \s)^2 (\pa\pa \s \pa \s \pa \s) \!-\! 12 \Box \s (\pa\pa \s)^2 (\pa \s)^2 \nn \\
& + 16 \Box \s (\pa\pa\pa \s \pa \s \pa \s \pa \s)- 24(\pa\pa \s \pa \s \pa \s)(\pa\pa \s)^2- 12 (\pa\pa \s)^2 (\pa \s)^4 + 12(\Box \s)^2 (\pa \s)^4\nn \\
&  - 20 \Box \s (\pa \s)^6 + 15(\pa \s)^8 \Bigl)\,.
\end{align}
In order to calculate the total action and specifically, to       read the value of $a^{*}_8$, we need to compute the above unknown  $\gamma_i$ coefficients. The first non-trivial terms appear at the 8th order of expansion. Similar to the previous cases one can show that the lower orders of the expansion are surface terms. Finally, after several by part integrations we find
\begin{align}
S_8 = &-\!\td{L}^7 \l_{eff}\!\!\int \! d^8x \Big(\! \frac{5}{128} (\pa \s)^8\! - \!\frac{5}{96} \Box\s(\pa \s)^6  \!- \!\frac{1073}{4032} (\Box \s)^2(\pa \s)^4 \! +\!  \frac{155}{576} \Box \s (\pa \s)^4 (\pa\pa \s)^2 \nn \\
& - \frac{299}{259} \,\Box \s(\pa \s)^2 \, (\pa\pa \s \pa \s \pa \s) + \frac{85}{72}\, (\pa \s)^2  (\pa\pa^a \s \pa\pa_a \s \pa \s \pa \s)+ \frac{1}{136}\, (\pa\pa \s \pa \s \pa \s)^2 \nn \\
& + \frac{1}{36}\, \Box \s\, (\pa\pa\pa \s \pa \s \pa \s \pa \s)  - \frac{17}{1008}\, \Box \s\,(\pa \s)^2  (\pa\pa \s)^2 + \frac{127}{2016}\, (\Box \s)^2\, (\pa\pa \s \pa \s \pa \s) \nn \\
&+ \frac{5}{1152}\Box^2 \s (\pa \s)^4  + \frac{11}{1344}(\Box \s)^3 (\pa \s)^2 \! -\! \frac{13}{336} (\pa\pa \s \pa \s \pa \s) (\pa\pa \s)^2+\frac{13}{36288} (\Box \s)^4 \nn \\
&- \frac{1}{672} (\Box \s)^2 (\pa\pa \s)^2 +\! \frac{1}{864} \Box \s \Box^2 \s(\pa \s)^2\! + \frac{1}{864} (\pa \s)^2 (\pa\pa\pa \s)^2 \!-\! \frac{5}{12096} ((\pa\pa \s)^2)^2 \nn\\
&- \frac{1}{432} (\pa\pa \s)^2 (\pa \s \pa \Box\s)\Big)\,.
\end{align}
By comparing this result with the dilaton action which is computed from equation \eqref{CompeleteS8}, one can read the unknown coefficients as
\bea
&&\g_1 = -\frac{4397}{114562} \g_5\,,\qquad 
\g_2 = \frac{7241}{16366} \g_5\,, \qquad 
\g_3 = -\frac{1012}{1169} \g_5\,,\qquad  \g_4 = -\frac{577}{1169} \g_5\,,  \nn \\
&&\g_5 = -\frac{167}{1866240} \td{L}^7 \l_{eff}\,,\qquad \g_6 = 0\,,\qquad \g_7 = 0\,,\qquad \g_8 = 0\,,\qquad \g_9 = 0\,,
\eea
and 
\be
a^*_8 = \frac{1}{6} \td{L}^7 \l_{eff}=\frac{\pi^{4} \td{L}^{5}}{6\k^2} \Bigl( \td{L}^2 - 16(9 \mathit{a}_1  + \mathit{a}_2 )\Bigr)\,.
\ee
In all the above calculations in even $d\leq 6$ dimension, the $a$-anomaly coefficients reduce to the GB results in \cite{Sinha:2012tc} when $a_1=a_2=0$.
Our results for $a^*_d$ confirm the general $d$ dimensional relation suggested by \cite{MyersSinha:2010tj}. For the GQC action in the Euclidean background the $a$-anomaly can be simply evaluated by computing the value of the bulk action on the AdS space 
\bea
\label{R2a*General}
a^*_d \!\!\!&=&\!\!\! \frac{\pi^{\frac{d}{2}} \td{L}^{d+1}}{d\, \G(\frac{d}{2})} \mathcal{L}_{bulk}\Big|_{AdS}\!\! = \frac{\pi^{\frac{d}{2}} \td{L}^{d-3}}{\k^2\G(\frac{d}{2})} \Bigl( \td{L}^2 \!-\! 2\big( \mathit{a}_1 \,d\,(d+1) + \mathit{a}_2 \,d + \mathit{a}_3 \,(d-1)\,(d-2)\big)\Bigr)\cr
\!\!\!&=&\!\!\! \frac{\pi^{\frac{d}{2}} {L}^{d-3}}{\k^2\G(\frac{d}{2})} \Bigl( {L}^2 - \frac{d+1}{2}\big( \mathit{a}_1 \,d\,(d+1) + \mathit{a}_2 \,d  +  \mathit{a}_3 \,(d-1)\,(d-2)\big)\Bigr)\,,
\eea
where in the last equality we have used the relation (\ref{LT}) and expanded it up to the first order of the  $a_i$ couplings. 

%%%%%%%%%%%%%%%%%%%%%%%%%%%%%%%%%%%%%%%%%%%%%
\section{Holographic RG flow in GQC gravity}
In the previous section, we established a holographic  $a$-theorem for the GQC gravity by finding the WZ action. In this section, we are going to study the holographic renormalization group (RG) flow of this theory in the presence of a matter field.
This RG flow  is  a function of the radial coordinate (RG scale) and the couplings of the theory ,\ie $a=a(r; a_1,a_2,a_3)$. We are interested in those functions, which are decreasing monotonically as we decrease the RG scale and are stationary at the UV/IR fixed points. The values of this function at these fixed points are given by $a^*_d$, the coefficients of the WZ action that we found in the previous section for even $d$ dimensions.
The $a$-theorem ensures that for any RG flow which connects the UV fixed point to the IR fixed point, $a_{UV}\geq a_{IR}$.

To study the holographic RG flow we begin with the gravitational action (\ref{R2action}) (in Minkowski signature) coupled to a matter field
\be
S=S_{bulk}+S_{Matter}\,.
\ee
We suppose that this matter field has various stationary points specifically at the UV and IR fixed points, therefore at these fixed points the vacuum solution of the equations of motion is AdS$_{d+1}$.

Now we consider a solution that   smoothly connects the two AdS space-times at the UV and IR fixed points (kink solution). This solution is a holographic representation of the renormalization group 
between the two dual boundary CFTs at the UV and IR fixed points
\be\label{metra}
ds^2=e^{2A(r)}\big(-dt^2+\sum_{i=1}^{d-1} dx_idx^i\big)+dr^2\,.
\ee

At the fixed points, this metric reproduces the AdS$_{d+1}$ solution, therefore at the UV fixed point $A(r)=r/L_{UV}$ and at IR, $A(r)=r/L_{IR}$. As the radial coordinate changes from $+\infty$ (UV fixed point) to $-\infty$ (IR fixed point), the RG flow which is a function of derivatives of $A(r)$, varies from $a_{UV}$ to $a_{IR}$.
By using the above geometry, the equations of motion in the presence of  the matter field can be written as
\begin{subequations}
\begin{align}
\label{EOMTrr}
\k^2 T^r{}_r&=-\frac{d}{2}\Big((d-1)(\frac{1}{L^2}-A'(r)^2)+{\k_2} A'(r)^4\nn \\
 &~~~+{\k_1}\big(2 d  A'(r)^2 A''(r)- A''(r)^2+2  A^{(3)}(r) A'(r) \big)\Big)\,, \\ 
\k^2 T^t{}_t=\k^2 T^i{}_i &=-\frac12 \Big(d(d-1)(\frac{1}{L^2}-A'(r)^2)-2 (d-1) A''(r) \nn \\
&~~~+{\k_2}\big(d  A'(r)^4+4 A'(r)^2 A''(r)\big)+{\k_1}\big(2 A^{(4)}(r) \nn \\ 
&~~~+3 d  A''(r)^2+4 d A^{(3)}(r) A'(r)+ 2d^2A'(r)^2 A''(r) \big) \Big)\,,\label{EOMTtt}
\end{align}
\end{subequations}
where ${T^{\mu}}_{\nu}$ is the energy-momentum tensor and $i=1,...,d-1$. These two equations only depend on two combinations of the three couplings of the theory
\be
\k_1=4d a_1+(d+1) a_2\,,\qquad 
\k_2=(d-3)\Big((d-1) (d-2) a_3 +d  ((d+1) a_1+a_2)\Big)\,.
\ee
\subsection{An ansatz for RG flow}
As it was mentioned at the end of the section two, the value of the $a$-anomaly is proportional to the value of the bulk Lagrangian computed on the AdS space-time \cite{MyersSinha:2010tj}. Away from the fixed points,  we expect that the value of the RG flow as a function of the energy, or holographically, the value of the $a$-anomaly as a function of the radial coordinate $r$, 
 is given by a function 
 $a(r)=a(A(r),A'(r),A''(r),...)$.
We define the following function as an ansatz for the holographic RG flow
\footnote{Such functionality comes into the mind by thinking to the Wilsonian approach in quantum field theory. It is possible that the RG flow, similar to its fixed points, is proportional or related to the effective Lagrangian, which is computed on the background solution. This  may contain higher curvature counter-terms for the gravitational field or the counter-terms for the matter field.}
\begin{align}\label{a}
a(r)=&\frac{\pi^{d/2}}{\k^2 \Gamma{(d/2)}}
\frac{1}{A'(r)^{d+1}} \nn 
\Big(A'(r)^2-\frac{2 \k_2 }{d-3}A'(r)^4+ \z_0A''(r) \\
&+\z_1  A'(r)^2  A''(r)+\z_2  A^{(3)}(r) A'(r)+\z_3 A''(r)^2 +\cdots\Big)\,.
\end{align}
This function reproduces the calculated values of $a^*$ in the section 2 when we insert $A(r)=r/\tilde{L}$. In other words, 
the coefficients of the first two terms are fixed by the value of $a^{*}$ at fixed points and there are no more terms such as $A'(r)^{n-d-1}$ for $n\geq 5$, as far as we study the GQC gravity.
We do not have any other constraint or condition to fix the remaining coefficients, which allows freedom on the profile of $a(r)$ and $a'(r)$ consequently. 
If we restrict ourselves to the above RG flow (ignoring the other possible terms in $...$), after the differentiation with respect to $r$, the $a'(r)$ has the same order of the derivatives as the equations of motion \eqref{EOMTrr} and \eqref{EOMTtt}.
As a result, the monotonicity of $a(r)$ depends on the behavior of $A'(r)$, $A''(r)$ and the energy-momentum components of the matter field, since we can get rid of $A^{(3)}(r)$ and $A^{(4)}(r)$ by using the equations of motion.

As it was discussed in section two,  $a^{*}_{UV}\geq a^{*}_{IR}$. In order to have a solution for equations of motion to support this condition, there will be some restrictions on the parameters of the theory.
By inserting the asymptotic values of the solution \ie $A(r)=r/\tilde{L}$, into the (\ref{a}), we must achieve  the values of $a^*$ in the UV/IR regions in equation (\ref{R2a*General})
\begin{equation}
a(r) \rightarrow \begin{cases}
    {{a}^{*}_{UV}}=\frac{\pi^{d/2}}{\k^2\G(d/2)} {\tilde{L}_{UV}}^{d-1}\big(1-\frac{2\k_2}{(d-3)\tilde{L}^2_{UV}}\big)\,, & ~ r\rightarrow +\infty\,,\\ \\
    {{a}^{*}_{IR}}=\frac{\pi^{d/2}}{\k^2\G(d/2)} {\tilde{L}_{IR}}^{d-1}\big(1-\frac{2\k_2}{(d-3)\tilde{L}^2_{IR}}\big)\,, & ~ r\rightarrow -\infty\,,
  \end{cases}
\end{equation}
where $\tilde{L}_{UV}$ and $\tilde{L}_{IR}$ are the effective radii of the AdS space in the $UV$ and $IR$ fixed points.
Since  $ {{a}^{*}_{UV}}\geq\ {{a}^{*}_{IR}}$,  by a simple algebraic analysis one may show  the following restrictions  on the value of $\k_2$
\begin{subnumcases}{}
 \tilde{L}_{UV}> \tilde{L}_{IR} ~~ \rightarrow ~~ -\infty<\frac{2\k_2}{d-3}<\tilde{L}_{UV}^2\,,  \label{co1} 
	\\ \nn\\
 \tilde{L}_{UV}< \tilde{L}_{IR} ~~ \rightarrow ~~ +\infty>\frac{2\k_2}{d-3}>\tilde{L}_{IR}^2\,.  \label{co2}
 \end{subnumcases}
These conditions only depend on the asymptotic behavior of the solutions of the equations of motion.
\subsection{The NEC and ANEC}
In theories of gravity in presence of matter fields, it is important to check whether the (Average) Null Energy Condition (A)NEC is valid or not. The NEC supposes that $\xi^\mu\xi^\nu T_{\mu\nu}\geq 0$, for all the null vectors $\xi^\mu$.
Although the NEC is valid for Einstein-Hilbert action, in the presence of the higher derivative terms with matter content, it does not have any explicit proof.
 For our propose, because of the symmetries of the metric \eqref{metra}, it would be enough to consider a null vector in the $(t,r)$ direction, then the null energy condition can be written as
\begin{align}\label{NEC}
\k^2({T^r}_r\!-\!{T^t}_{t})=&\big(1-d+2\k_2A'(r)^2\big)A''(r)\nn \\
&+\k_1\big(2d A''(r)^2+d A'(r) A^{(3)}(r)+A^{(4)}(r)\big)\geq 0\,.
\end{align}
In general it is impossible to find regions in the space of the couplings where this inequality holds without knowing the exact functionality of the $A(r)$. 

On the other hand, ANEC is a weaker condition. It states that along a complete null curve, here from the UV to IR fixed point, the negative energy fluctuations cancel by positive energy fluctuations. In another word 
\be\label{ANEC}
\int_{-\infty}^{+\infty}\k^2({T^r}_r\!-\!{T^t}_{t})dr \geq 0\,.
\ee
A holographic proof of the ANEC is presented in \cite{Kelly:2014mra}. Also for interacting theories, it has been shown that the ANEC is coming from the micro-causality in unitary quantum field theories \cite{Hartman:2016lgu}.
Inserting the right hand side of \eqref{NEC} into \eqref{ANEC} we conclude that
\be\label{IANEC}
(1-d)\big(\frac{1}{\tilde{L}_{UV}}-\frac{1}{\tilde{L}_{IR}}\big)+\frac23 \k_2 \big(\frac{1}{\tilde{L}_{UV}^3}-\frac{1}{\tilde{L}_{IR}^3}\big)+\k_1 d \int_{-\infty}^{+\infty} A''(r)^2 dr \geq 0\,.
\ee
For Einstein-Hilbert action $(\k_1=\k_2=0)$ the above inequality reduces to $\tilde{L}_{UV}\geq \tilde{L}_{IR}$. However, in presence of the higher curvature terms we will find a linear condition for parameter space of $\k_1$ and $\k_2$.
Although the value of ANEC depends on the choice of $A(r)$ or $A''(r)$ but this dependence appears as a positive coefficient in \eqref{IANEC} and therefore, ANEC has a universal behavior for all the possible solutions of the equations of motion.

\subsection{A toy model}
To have an overall view and to examine the behavior of the RG flow of (\ref{a}), as a simple toy model, suppose the following function is a solution of the equations of motion in the presence of a proper matter field
\begin{align}\label{solut}
A(r)=\big(\frac{1}{\tilde{L}}+A B\big)r + A\, Ln\big(cosh(B\,r)\big)\,.
\end{align}
Similar smooth functions have been found as the solution of equations of motion in specific theories of matter coupled to gravity, for example, see \cite{Freedman:2003ax}. For instance, by adding a super-potential to the Einstein-Hilbert action one may find an exact solution for $A(r)$ and matter fields. Here in the presence of the higher curvature terms, we consider such a super-potential exists (or even simpler, we can just consider a massless scalar field with a kinetic term) and the equations of motion \eqref{EOMTrr} and \eqref{EOMTtt} support the above solution. 

In this model, we have  freedom to choose the sign of $B$, so we suppose that $B<0$. The ansatz of \eqref{solut}  asymptotically admits  the AdS solution $A(r)= r/\tilde{L}_\infty$ so that
\begin{align}
\tilde{L}_\infty = \begin{cases}
    {\tilde{L}_{UV}}={\tilde{L}}\,, & r\rightarrow +\infty\,,\\ \\
    {\tilde{L}_{IR}}=\frac{\tilde{L}}{1+2 A B \tilde{L}}\,, & r\rightarrow -\infty\,.
  \end{cases}
\end{align}
As we see, if $A>0$ then $ {\tilde{L}_{UV}}< {\tilde{L}_{IR}}$ and if $A<0$ then $ {\tilde{L}_{UV}}>{\tilde{L}_{IR}}$. 
The value of $\tilde{L}_\infty$ in the UV/IR fixed points is related to the asymptotic value of the energy-momentum tensor components via the following quadratic equations
\be
\k^2{T^r}_r\Big|_{r\rightarrow\pm\infty}=\k^2{T^{t}}_t\Big|_{r\rightarrow\pm\infty}=-\frac{d}{2\tilde{L}^4_{\infty}}\Big(\k_2-(d-1)\tilde{L}^2_{\infty} \big(1-\frac{\tilde{L}^2_{\infty}}{L^2}\big)\Big)\,.
\ee
%The greater(smaller) root of this equation is %$\tilde{L}_{UV}$($\tilde{L}_{IR}$).

\begin{figure}[!htbp]
\centering
\includegraphics[width=0.6\textwidth]{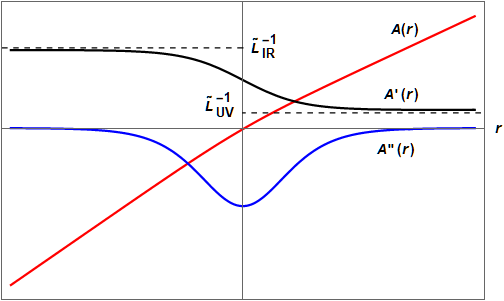}
\caption{\footnotesize{The red, black and blue curves represent the behavior of the solution (\ref{solut}) and its first and second derivatives, $A,B<0$. This solution asymptotically goes to the AdS solution in the UV/IR regions. }}\label{fig1}
\end{figure}

As an example, the behavior of the solution (\ref{solut}) and its first and second derivatives are depicted in figure \ref{fig1} for $A<0$ and $B<0$.

In the following subsections, we will need to know the behavior of the $a(r)$. Specifically, we are interested in  its monotonic behavior. As a result, one should examine the behavior of the $a'(r)$ under the various values of $\z_i$ coefficients in \eqref{a}. This provides various paths in the RG flow depending on the choice of $\z_i$ coefficients, although all paths asymptotically have the same behavior at $r\rightarrow \pm\infty$ where $a'(r)\rightarrow 0$. 

\subsubsection{Regions of NEC and ANEC}
For our toy model, we can examine both the NEC and ANEC regions of validity. By inserting the solution (\ref{solut}) into the \eqref{NEC}, the null energy condition would be
\begin{align}
\k^2({T^r}_r\!-\!{T^t}_{t})=&\frac{x\l^2}{A^3 \tilde{L}^4(1+x)^4}
\Big(  \k_1\l^2(1 + (x-4 ) x)  -   \k_1 \l A d ( x^2 +(x-3)x\l-1) \nn \\
&+ 2 A^2\k_2 (1 + (1+\l)x)^2-A^2( d-1) \tilde{L}^2 (1 + x)^2 \Big)\geq 0\,,
\end{align}
where for short notation we have defined $e^{2 B r}=x>0$ and  $2AB\tilde{L}=\l$. The sign of the overall coefficient on the right hand side only depends  on the sign of $A$, therefore the expression in the parenthesis above, which is a quadratic polynomial of $x$, must be positive/negative everywhere in the interval of $0<x<+\infty$ when $A>0$ or $A<0$. 

\begin{figure}[!htbp]
\centering
\includegraphics[width=0.6\textwidth]{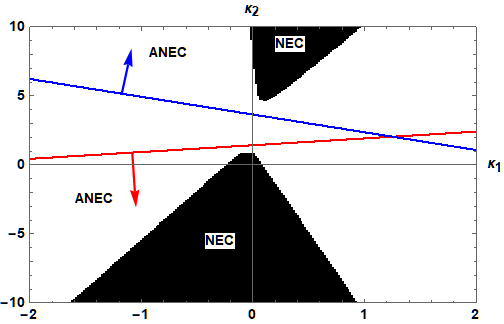}
\caption{\footnotesize{The regions of  validity for NEC and ANEC. The size of the wedges depend on the values of $A, B$ and $\tilde{L}$. This graph is depicted for the special values, $d=4$, $A=\pm0.1$, $B=-2$ and $\tilde{L}=1.2$. The ANEC holds below the red line for $A<0$ and above the blue line for $A>0$.}}\label{NECP}
\end{figure}

By a simple numerical analysis, we can find the regions in $(\k_1,\k_2)$ space where the null energy condition holds, see figure \ref{NECP}. In this figure the upper wedge corresponds to the $A>0$ and the lower wedge to the $A<0$. The size of wedges depends on the values of $A, B$ and $\tilde{L}$. We have also considered  (\ref{co1}) and (\ref{co2}) conditions  in the drawing of the figure \ref{NECP}.

On the other hand, the value of ANEC can be computed as follow
\bea
\int_{-\infty}^{+\infty}\!\!\!\!\k^2({T^r}_r\!-\!{T^t}_{t})dr=\frac{-\l}{6A\tilde{L}^3}\big(d \l^2 \k_1 +4 A (\l^2+3\l+3)\k_2-6A(d-1)\tilde{L}^2\big)\geq 0\,.
\eea
For  the positive values of $A$ the ANEC exists above the line of $d \l^2 \k_1 +4 A (\l^2+3\l+3)\k_2-6A(d-1)\tilde{L}^2=0$ in $(\k_1,\k_2)$ space and for the negative $A$, below that line (see figure \ref{NECP} for a specific choice of parameters).
 
\subsubsection{A monotonically decreasing RG flow}
As it was mentioned in the introduction, we expect that the RG flow monotonically decreases because in the Wilsonian approach by integrating out the high energy modes, the number of degrees of freedom decreases. This means that we must look for the RG flows that $a'(r)\geq 0$.
Generally, it is hard to find a set of specific values for $\z_i$ coefficients in (\ref{a}) or a condition on the matter field, such as the (A)NEC, to prove $a'(r)\geq 0$.
To investigate the behavior of the RG flow we study two examples by fixing the free parameters $\z_i$ in the ansatz (\ref{a}). After that, we discuss about a more general case.
\subsubsection*{Example 1}
\begin{figure}[!htbp]
\centering
\includegraphics[width=0.6\textwidth]{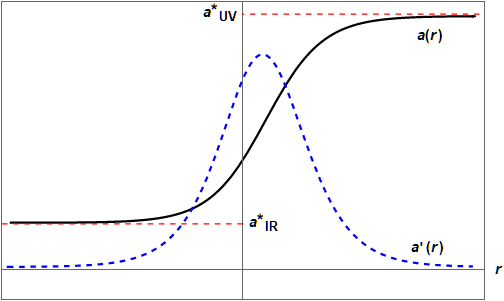}
\caption{\footnotesize{Holographic RG flow in the example 1.}}\label{fig2}
\end{figure}

In the simplest example,  let's turn off all the extra terms in the second line of the equation \eqref{a} \ie $\z_0=\z_1=\z_2=\z_3=0$ then
\begin{align}\label{apex1}
a'(r)&=\frac{\pi^{d/2}}{\k^2 \Gamma{(d/2)}}\frac{A''(r)}{A'(r)^{d}}\Big\{ 2 \k_2 A'(r)^2-d+1\Big\} \nn\\
&=\frac{-\pi^{d/2}  \tilde{L}^{d-4}}{A\k^2 \Gamma{(d/2)}} \frac{\l^2x(1+x)^{d-4}}{(1+(1+\l)x)^d} \Big((d-1)\tilde{L}^2(1+x)^2-2\k_2(1+(1+\l)x)^2\Big)\,.
\end{align}
This equation clearly is independent of the $\k_1$.  To check whether \eqref{apex1} is positive or not it is sufficient to search all the possible roots of the above equation. Depending on the sign of $A$, the overall coefficient is positive/negative therefore the expressions inside the parenthesis must be positive/negative everywhere in $x>0$. A simple analysis of the quadratic polynomials shows that, to have a monotonically decreasing RG flow for every value of $\k_1$, it requires the $\k_2$  takes the following values 
\begin{align}
\begin{cases}
    -\infty<\k_2< \frac{(d-1)\tilde{L}^2}{2(1+\l)^2}\,, & A< 0\,,\\ \\
   +\infty>\k_2>\frac{(d-1)\tilde{L}^2}{2(1+\l)^2}\,, & A>0\,.
  \end{cases}
\end{align}
 These regions always have an overlap with the region of validity of the (A)NEC.
A general behavior of $a(r)$ and $a'(r)$ is sketched in figure \ref{fig2}. 

\subsubsection*{Example 2}
Another interesting example is the specific choice of $\z_0=0,\, \z_1=d\k_1,\, \z_2=\k_1$ and $\z_3=\frac{d}{2}\k_1$ in the equation (\ref{a}). With the help of the null energy condition \eqref{NEC} and equations of motion \eqref{EOMTrr} and \eqref{EOMTtt} we can write $a'(r)$ as follow
\begin{align}\label{apex2}
a'(r)&=\frac{\pi^{d/2}}{\k^2 \Gamma{(d/2)}}\frac{1}{A'(r)^{d}}\Big\{\Delta\mathcal{T}-d(d+1)\k_1 A''(r)^2\big(1+\frac{A''(r)}{2A'(r)^2}\big)\Big\}\,. 
\end{align}
$\bullet$ $\k_1=0:$
To have a monotonically decreasing $a$-function, one possible choice  is  $\k_1=0$,  together with a matter field that satisfies the NEC, $\Delta\mathcal{T}=\k^2({T^{r}}_r-{T^{t}}_t)>0$, (this coincides with the example 1 for $\k_1=0$). The pure gravitational part of the theory restricts to a specific type of the quadratic curvature Lagrangian
\be
\mathcal{L}_{bulk}^{Gr}=\mathit{a}_1 (R^2-\frac{4d}{d+1} R_{\m\n}^2) + \mathit{a}_3 (R^2 - 4 R_{\m\n}^2+R_{\m\n\r\s}^2)\,.
\ee
This result agrees  with the result of \cite{BhattacharjeeSinha:2015qaa} which has been done in $d=4$ by another approach. Also the condition $\k_1=0$ 
 avoids the propagation of the scalar degrees of freedom of the graviton modes in AdS background, see for example \cite{Deser:2011xc, Ghodsi:2017iee}.
\\
$\bullet$ $\k_1\neq 0:$
By inserting the solution (\ref{solut}) into the equation (\ref{apex2}) we find the following expression 
\begin{align}
a'\big(x(r)\big)=&-\frac{\pi^{d/2}}{\k^2 \Gamma{(d/2)}}\frac{\l^2 \tilde{L}^{d-4} }{A^3}\frac{ x (1+x)^{d-4}}{ (1+(1+\l)x)^{d+2}} \mathcal{F}(x)\,, \\
\qquad \mathcal{F}(x)=&A^2 \big(1 + (1+\l) x \big)^2 \Big((d-1) \tilde{L}^2 (1 + x)^2 - 2\k_2 (1 + (1+\l)x)^2 \Big) \nn \\
&
+\k_1  d  A \l \big(1 + (1+ \l)x\big)^2 \Big(x^2 +  (x+d\!-\!2)x \l\!-\!1\Big) \!-\! 
 \k_1 \l^2 \Big(1+ 2x(\l\!-\!1) \nn \\
&+ x^2 \big(  (\l+1)^2 x^2 \!-\! 2  (2\l+1)(\l+1)x \!-\!\frac12 (d-1)(d+2) \l^2-6(\l+1)\big)\Big) \,.\nn
\end{align}
Associated to the sign of $A$, the coefficient of $\mathcal{F}(x)$ is a positive/negative function. Moreover this coefficient asymptotically goes to zero on both UV $(x\rightarrow +\infty)$ and IR $(x\rightarrow 0)$ sides and it  has just one extremum point. 
The $\mathcal{F}(x)$ itself is a fourth order polynomial of $x$, therefore if we demand a monotonically decreasing function of $a(r)$,  we must find conditions which $\mathcal{F}(x)$ is negative/positive for all the values of $x>0$. It means that this function should not have any root in this interval. 

\begin{figure}[!htbp]
\centering
\includegraphics[width=0.6\textwidth]{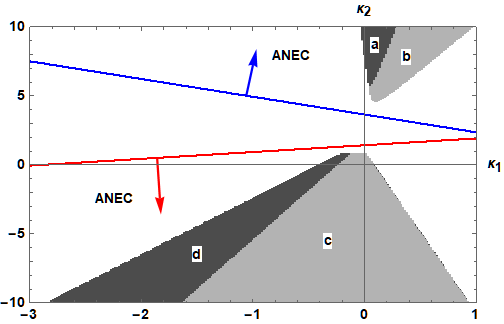}
\caption{\footnotesize{Monotonicity, NEC and ANEC regions. This graph is depicted for special values, $d=4$, $A=\pm 0.1$, $B=-2$ and $\tilde{L}=1.2$, but for every value of these parameters in the allowed regions we always observe the same behavior.}}\label{k1-k2}
\end{figure}

Analytic analysis of the fourth order polynomials gives a set of very complicated inequalities, instead, we present a numerical analysis in this paper. Our results hold in every dimension and the graphs are generally the same when we fix the values of $A, B, \tilde{L}$ or $\l$.
A numerical survey in the space of the couplings $(\k_1,\k_2)$ shows the region where $\mathcal{F}(x)$ is positive/negative for all the values of $x>0$, see figure \ref{k1-k2}. 

By defining,
$\mathcal{M}=\big\{\!\!$ set of points where $a'(r)>0\big\}$
and
$\mathcal{N}=\big\{\!\!$ set of points where the NEC holds$\big\}$,
we would have the following  $a, b, c$ and $d$ regions in  figure \ref{k1-k2} 
\begin{align}
\begin{cases}
A>0:
\, a \in \mathcal{M}\,, \quad a\cup b \in \mathcal{N},\nn \\ \\
A<0:
\, c\cup d \in \mathcal{M}\,, \quad c\in \mathcal{N}.\nn
\end{cases}
\end{align}
As we see, all points in the region $\mathcal{M}$ lay in the allowed regions by ANEC. But this is not correct for the NEC region.
\subsubsection*{A general analysis}
\begin{figure}[!htbp]
\centering
\begin{subfigure}{0.49\textwidth}
\includegraphics[width=1\textwidth]{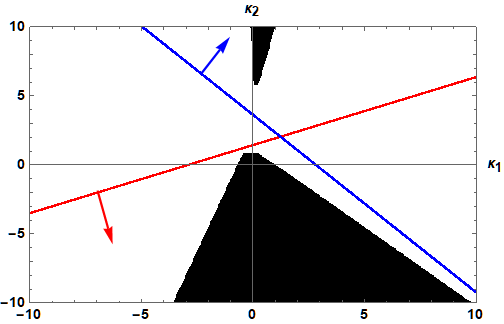}
\caption{\footnotesize{$\z_0=\k_1$}}\label{x0z1}
\end{subfigure}
\begin{subfigure}{0.49\textwidth}
\centering
\includegraphics[width=1\textwidth]{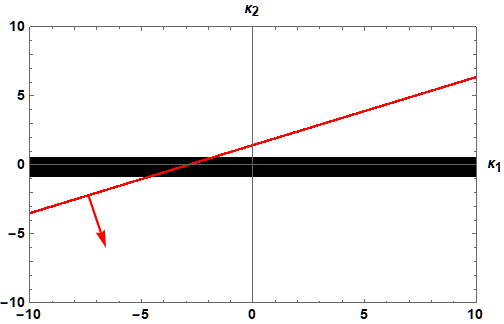}
\caption{\footnotesize{$\z_0=\k_2$}}\label{x0z2}
\end{subfigure}
\begin{subfigure}{0.49\textwidth}
\centering
\includegraphics[width=1\textwidth]{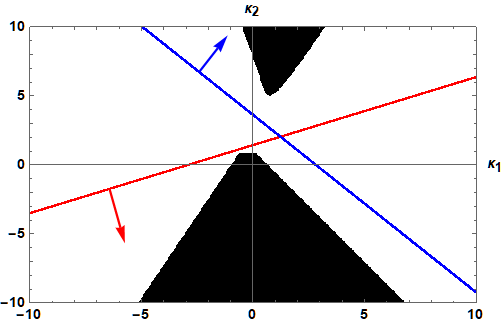}
\caption{\footnotesize{$\z_1=\k_1$}}\label{x1z1}
\end{subfigure}
\begin{subfigure}{0.49\textwidth}
\centering
\includegraphics[width=1\textwidth]{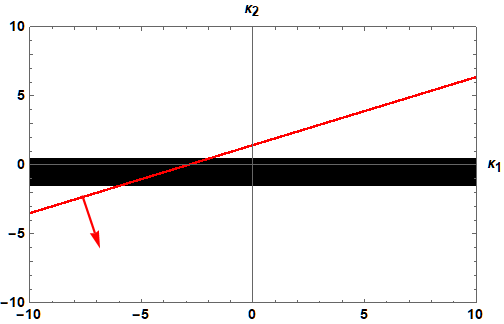}
\caption{\footnotesize{$\z_1=\k_2$}}\label{x1z2}
\end{subfigure}
\begin{subfigure}{0.49\textwidth}
\centering
\includegraphics[width=1\textwidth]{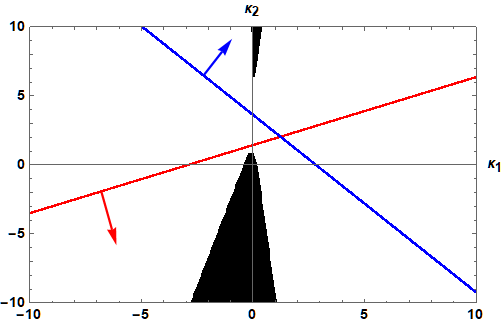}
\caption{\footnotesize{$\z_2=\k_1$}}\label{x2z1}
\end{subfigure}
\begin{subfigure}{0.49\textwidth}
\centering
\includegraphics[width=1\textwidth]{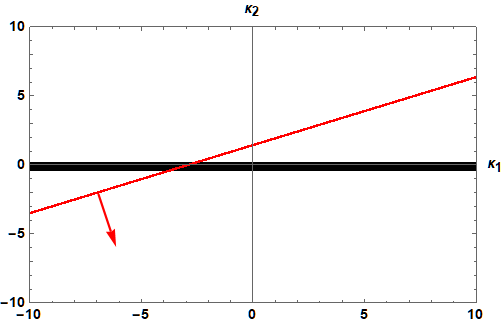}
\caption{\footnotesize{$\z_2=\k_2$}}\label{x2z2}
\end{subfigure}
\begin{subfigure}{0.49\textwidth}
\centering
\includegraphics[width=1\textwidth]{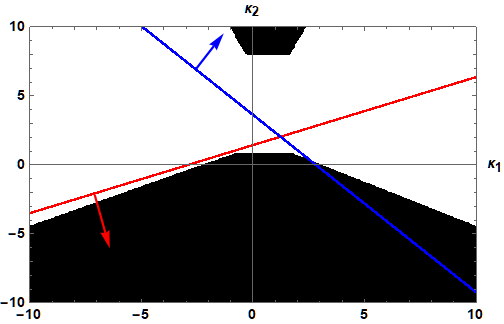}
\caption{\footnotesize{$\z_3=\k_1$}}\label{x3z1}
\end{subfigure}
\begin{subfigure}{0.49\textwidth}
\centering
\includegraphics[width=1\textwidth]{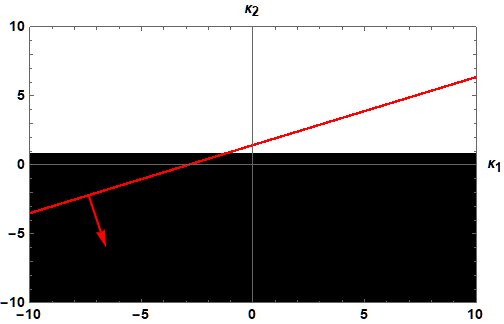}
\caption{\footnotesize{$\z_3=\k_2$}}\label{x3z2}
\end{subfigure}
\caption{\footnotesize{All figures are depicted for special values, $d=4$, $A=\pm 0.1$, $B=-2$ and $\tilde{L}=1.2$.}}\label{gpic}
\end{figure}

It would be interesting to look at the general form of  the RG flow. Again we insert the solution (\ref{solut}) into the general form of $a(r)$ in equation (\ref{a}). By a differentiation with respect to $r$ we have
\begin{align}
a'\big(x(r)\big)=&-\frac{\pi^{d/2}}{\k^2 \Gamma{(d/2)}}\frac{\l^2 \tilde{L}^{d-4} }{A^3}\frac{ x (1+x)^{d-4}}{ (1+(1+\l)x)^{d+2}} \mathcal{G}(x)\,, 
\end{align}
where
\begin{align}
\mathcal{G}(x)=&  A^2\big(1 + (1+\l)x\big)^2 \Big((d-1) \tilde{L}^2 (1 + x)^2 \! - \!2 \k_2 \big(1 + (1+ \l)x\big)^2\Big)\! + A \l \Big(\z_1 \big(1+ (1 \nn \\
&+\l)x\big)^2 \big((1 + \l) x^2  + (d-2)\l x -1 \big) + \z_0\tilde{L}^2 (1 + x)^2  \big(  (1+ \l)x^2 + d \l x -1 
\big)\Big)\nn \\
&- \l^2 \Big(\z_2 \big(1 + (1+ \l)x\big) \big( (1 + \l)x^3  + 
((d-4) \l-3) x^2 \! -\! ((d-1) \l+3)x+1\big) \nn \\
&  - \z_3 \l x \big( 2 (1 + \l) x^2 + (d -1) \l x -2 \big)\Big)\,.
\end{align}
The analysis is similar to the analysis of example 2, because $\mathcal{G}(x)$ is also a fourth order polynomial of $x$. For simplicity let's suppose  that the  $\z_i$ coefficients are linear combinations of the $\k_1$ and $\k_2$, \ie $\z_i=a_i \k_1+b_i \k_2$ for $i=0, 1, 2, 3$. 
To follow the effect of each term in \eqref{a} individually, we separate each $\z_i$  by setting all the other $\z$'s equal to zero, see the figures \ref{x0z1} to \ref{x3z2}.

 In all the left hand side figures \ref{x0z1}, \ref{x1z1}, \ref{x2z1} and \ref{x3z1} we have fixed all $b_i=0$. The lower wedges belong to $A<0$ and the upper ones are for $A>0$. 
On the other hand in all the right hand side figures, \ref{x0z2}, \ref{x1z2}, \ref{x2z2} and \ref{x3z2}  we have fixed all $a_i=0$. These strips (half plane) belong to $A<0$, in fact,  $A>0$ has not any allowed region. For all cases, we have drawn the regions of validity of ANEC. Again these behaviors are general in every dimension and independent of the values of $A, B$ and $\tilde{L}$.
 
From these figures, we can conclude the following general results:
\begin{itemize}
	\item If all $a_i=0$ then the allowed region would be a strip (half plane) in $(\k_1,\k_2)$ plane. The width of this strip depends on the choice of the coefficients as well as the values of $A, B$ and $\tilde{L}$. This case is similar to example 1.
	\item If at least one of the $a_i\neq 0$ then there will be two wedge-like regions in the $(\k_1,\k_2)$ plane similar to the example 2.
\end{itemize}

%%%%%%%%%%%%%%%%%%%%%%%%%%%%%%%%%%%%%%%%%%%%%%%%%%%%%%
\section{Summary and Discussion}
This paper is divided into two main parts, we first holographically show the existence of the $a$-theorem for even dimensional conformal field theories which are dual to the AdS space in general quadratic curvature gravity. In the second part, we discuss on the holographic RG flow between two CFT's at the UV and IR fixed points.

In section two we generalize the method in reference \cite{Sinha:2012tc}, which in the context of the gauge/gravity correspondence we find the effective dilaton action corresponding to the spontaneously broken conformal symmetry in even dimensions. At the first step, we need the GH terms and counter-terms corresponding to the bulk action of \eqref{R2actionWith a3}. 

The GH terms for Einstein-Hilbert and Gauss-Bonnet terms are known but with the standard method of variation, one cannot find a proper GH term for general quadratic curvature terms. We do this by computing the effective GH term on a maximally symmetric AdS space \cite{Cremonini:2009ih}. The final result  \eqref{r2 gh} is a GH term for EH action but with an effective coefficient. The total GH surface terms are the sum of the GH terms in \eqref{gb gh} and  \eqref{r2 gh}.

The counter-terms for general quadratic curvature gravity already are computed by various approaches up to the quadratic boundary curvatures \cite{Cremonini:2009ih}. We use the algorithm in  \cite{Kraus:1999di} to compute these counter-terms up to the cubic curvature terms which are needed to study the conformal field theories with dimensions $d\leq 8$. The total counter-terms are the sum of (\ref{LctGB}), (\ref{Lct0}) and (\ref{Lct121}) - (\ref{Lct123}).

After finding all the necessary Lagrangians, following to   \cite{Sinha:2012tc}, we introduce a radial cut-off as a scalar function of the boundary variables. By using the induced metric on this cut-off surface and computing the bulk and surface terms we find the WZ action of the dilaton field in $d=2, 4, 6, 8$.
In all of these dimensions, the coefficient of the WZ action is the value of $a$-anomaly and agrees with the known relation of \eqref{R2a*General} for $a^*_d$ in term of the bulk action computed on the AdS background. Moreover, in $d=8$ as well as the WZ terms we can find some non-vanishing Weyl invariant terms which already are introduced in \cite{Elvang:2012yc}.

The existence of this WZ action holographically shows that the $a$-theorem exists for conformal field theory dual to the AdS space in GQC gravity in even $d$ dimensions. 

In section three, we study the holographic renormalization group flow in GQC gravity in the presence of a matter field.
We try to find those RG flows that monotonically are decreasing as we decrease the RG scale and are stationary at the UV/IR fixed points. The value of the RG flow at these fixed points is given by $a^*_d$ that we found in section two.
The $a$-theorem ensures that for any RG flow which connects the UV fixed point to the IR fixed point, $a_{UV}\geq a_{IR}$.

In this section we use the ansatz of \eqref{a} for RG flow which is constructed from the warped factor of the kink solution \eqref{metra}. This kink solution is interpolating between the two AdS solutions in the UV/IR fixed points. The $a$-theorem makes restrictions \eqref{co1} and \eqref{co2} on the value of the couplings.  

To study the RG flow we need to know the exact form of the kink solution from equations of motion. In the presence of a matter field, this is not a simple job, instead, we use the toy model of \eqref{solut} which has all the properties we need. 
Meanwhile, since we are studying the gravity in the presence of a matter field, it is important to check the regions of the validity of the (average) null energy condition. We have presented a numerical sample of our results in figure \ref{NECP}. Our numerical analysis shows that the NEC in this toy model allows not all the possible values of the couplings. On the other hand, the  ANEC as a weaker condition provides a wider region. We expect that by imposing the ANEC the dual quantum field theories do not suffer from the negative energy fluctuations as proved by \cite{Kelly:2014mra}. 

Finally, we have studied the RG flow \eqref{a} in two examples by fixing the free parameters in \eqref{a}. We observe a general behavior for the allowed region where the monotonically decreasing RG flow exists. The numerical results are summarized in figures \ref{k1-k2} and \ref{gpic}.
We show that the regions of monotonically decreasing RG flow may or may not have overlap with the regions where the null energy condition holds, for example, see figure \ref{k1-k2}.

The analysis of figure  \ref{gpic} suggests that, if we demand the ANEC together with a monotonically decreasing RG flow then the unknown $\z_i$ coefficients in the RG flow of \eqref{a} must be just a function of $\k_1$ and not $\k_2$. Therefore we believe that example 2 is a good description for the RG flow and changing the numerical coefficients of $\z_i$ coefficients does not alter the whole picture. 
%%%%%%%%%%%%%%%%%%%%%%%%%%%%%%%%%%%%%%%%%%%

\appendix
\section{Boundary curvatures}
We first compute the Riemann tensor by using the induced metric (\ref{BoundaryMetric}) and then expand the results in term of the derivative of $\s$. Up to the 8 derivatives we find
\begin{align}
\label{Riemann}
\mathcal{R}_{abcd} 
 &=  \td{L}^2\big(e^{-2\s} \big(\d_{ac} (\pa_{b}\s \pa_{d}\s + \pa_{d}\pa_{b}\s) + \d_{bd} (\pa_{a}\s \pa_{c}\s + \pa_{c}\pa_{a}\s- \d_{ac} (\pa \s)^2) \big)\nn\\
&+ \big(1 - e^{2\s} (\pa \s)^2+e^{4\s} (\pa \s)^4\big) \big( \d_{bc} \pa_{d}\pa_{a}\s (\pa \s)^2 + \pa_{c}\pa_{a}\s (\pa_{d}\pa_{b}\s - \d_{bd} (\pa \s)^2)\nn\\
& + \d_{ac} \d_{bd} (\pa \s)^4  + 2 \pa_{b}\s \pa_{d}\s (\pa_{c}\pa_{a}\s - \d_{ac} (\pa \s)^2) -   2 \pa_{b}\s\pa_{c}\s (\pa_{d}\pa_{a}\s -  \d_{ad} (\pa \s)^2)\big)\big) \nn \\
&-(a\leftrightarrow b)\,.
\end{align}
By a proper contraction of indices, the  Ricci tensor is
\begin{align}
\label{RicciT}
\mathcal{R}_{ab} &=  (d-2)\big(\pa_a \s \pa_b \s + \pa_a\pa_b \s) + \big(\pa^2 \s - (d-2)(\pa \s)^2 \big) \d_{ab}-  e^{2\s} \bigl[ 2 \pa_a \s \pa_b \s \big((d-2)(\pa \s)^2 \nn \\
& - \pa^2 \s \big)+ \pa_a\pa_b \s \big( (d-3)(\pa \s)^2 - \pa^2 \s \big) + \pa_{c}\pa_{b}\s \pa^{c}\pa_{a}\s +\pa_{b}\s \pa_{c}\pa_{a}\s \pa^{c}\s+ \pa_{a}\s \pa_{c}\pa_{b}\s \pa^{c}\s \nn \\
& 
-\! \d_{ab} \big( (d-3)(\pa \s)^4\! -\! (\pa \s)^2 \pa^2 \s \!-\! (\pa_a\pa_b \s)^2 \big)\bigr]+  e^{4\s} \bigl[  \pa_a \s \pa_b \s \big( (2d\! -\! 5) (\pa \s)^4 \!-\! 2 (\pa \s)^2 \pa^2 \s \nn \\
&  - 2  \pa_{c}\pa_{d}\s \pa^{c}\s \pa^{d}\s \big)+ \pa_a\pa_b \s \big( (d -4) (\pa \s)^4  - (\pa \s)^2 \pa^2 \s - \pa_{c}\pa_{d}\s \pa^{c}\s \pa^{d}\s \big)+ 2  ( \pa_{b}\s \pa_{d}\pa_{a}\s \nn \\
& + \pa_{a}\s \pa_{d}\pa_{b}\s )\, \pa^d\s\, (\pa \s)^2 + \pa^{c}\s \pa_{d}\pa_{b}\s \,\big(\pa_{c}\pa_{a}\s \pa^{d}\s  + \pa_{c}\s \pa^{d}\pa_{a}\s \big)
\!-\!  \big(  2 (\pa \s)^2 \,\pa_{c}\pa_{d}\s\, \pa^{c}\s \,\pa^{d}\s \nn \\
& -(d-4) (\pa \s)^6 + (\pa \s)^4  \pa^2 \s\big) \d_{ab} \bigr]
- e^{6\s} (\pa \s)^2 \big[ 2 \pa_a \s \pa_b \s \big( (d - 3) (\pa \s)^4 - 2 \pa_{c}\pa_{d}\s \pa^{c}\s \pa^{d}\s \nn \\
&- (\pa \s)^2 \pa^2  \s\big)
   + 3 \pa^d\s (\pa \s)^2  \big( \pa_{b}\s \pa_{d}\pa_{a}\s +\pa_{a}\s \pa_{d}\pa_{b}\s \big)+ \pa_a\pa_b \s \big( (d -5) (\pa \s)^4 \!-\! (\pa \s)^2 \pa^2 \s\nn \\
& - 2 \pa_{c}\pa_{d}\s \pa^{c}\s \pa^{d}\s \big)+ \big(2\,\pa_{c}\pa_{a}\s \pa^{d}\s + \pa_{c}\s \pa^{d}\pa_{a}\s \big) \pa^{c}\s \pa_{d}\pa_{b}\s - \big( (d-5) (\pa \s)^6\! -\! (\pa \s)^4  \pa^2 \s\nn \\
&  - 3 (\pa \s)^2 \pa_{c}\pa_{d}\s \pa^{c}\s \pa^{d}\s \big) \d_{ab}
\big]\,,
\end{align}
and the Ricci scalar is
\begin{align}
\label{RicciS}
\mathcal{R} &= \frac{1  }{\td{L}^2}\big((d-1)e^{2\s}\bigl( 2 \pa^2 \s \!-\! (d\!-\!2) (\pa \s)^2 \bigr) \! + \! e^{4\s}\bigl( (d-1)(d-4) (\pa \s)^4\! -\! 2(d-2) (\pa \s)^2 \pa^2 \s  \nn \\
&  + (\pa^2 \s)^2\!-\! (\pa_a \pa_b \s)^2 \!-\! 2 \pa_a \pa_b \s \pa^a \s \pa^b \s \bigr) + e^{6\s}\big( (d-1)(d-6) (\pa \s)^6 \!-\! 2 (d-3) (\pa \s)^4 \pa^2 \s \nn \\
& + 2( \pa_{a}\pa_{b}\s \pa^2 \s -  \pa_{b}\pa_{c}\s \pa^{c}\pa_{a}\s) \pa^{a}\s \pa^{b}\s \! - (\pa \s)^2 ( 4  d \pa_{a}\pa_{b}\s \pa^{a}\s \pa^{b}\s+ (\pa_a \pa_b \s)^2 -(\pa^2 \s)^2)\big)\nn \\
& + e^{8\s} (\pa \s)^2 \big( (d-1)(d-8) (\pa \s)^6 \!-\! 2  (d-4) (\pa \s)^4 \pa^2 \s + (\pa \s)^2 (\pa^2 \s)^2 - (\pa \s)^2 (\pa_a \pa_b \s)^2 \nn \\
& - 6  d (\pa \s)^2 \pa_{a}\pa_{b}\s \pa^{a}\s \pa^{b}\s + 4 \pa_{a}\pa_{b}\s \pa^{a}\s \pa^{b}\s \pa^2 \s - 4 \pa_{b}\pa_{c}\s \pa^{c}\pa_{a}\s \pa^{a}\s \pa^{b}\s \big)\big)\,.
\end{align}
Providing the intrinsic curvatures, the next step would be calculating the proper extrinsic curvature from \eqref{BoundaryMetric}. The space-like vector normal to the boundary surface is
\be
\label{NormalVector}
n_\m = (n_z , n_a) = \frac{\td{L}}{z \sqrt{1+ (\pa z)^2}} (-1 ,\,  \pa_a z)\,.
\ee
By definition, the extrinsic curvature tensor $K_{ab}$ is equal to
\be
K_{ab} = \nabla_\b n_\a e^\a_a e^\b_b\,,
\ee
where $e^\a_a$, the projector operator (pullback tensor) of the bulk to the boundary, is defined as $e^\a_a = \frac{\pa x^\a}{\pa x^a}$.
The final results for  $K_{ab}$ and $K$ up to the 8 derivatives  are
\begin{align}
\label{Kab}
K_{ab} =\td{L}\Big( & e^{-2 \s} \d_{ab} +   2  \pa_a \s \pa_b \s + \pa_a \pa_b \s - \hlf (\pa \s)^2 \d_{ab} -  \frac{ e^{2 \s}}{8}  \bigl(  8  \pa_a \s \pa_b \s + 4  \pa_a \pa_b \s\nn\\
&  - 3  (\pa \s)^2 \d_{ab} \bigr) (\pa \s)^2+   \frac{ e^{4 \s}}{16}  \bigl(  12  \pa_a \s \pa_b \s + 6  \pa_a \pa_b \s - 5  (\pa \s)^2 \d_{ab} \bigr)(\pa \s)^4 \nn \\
&-   \frac{5 e^{6 \s}}{128}  \bigl(  16  \pa_a \s \pa_b \s  + 8  \pa_a \pa_b \s - 7  (\pa \s)^2 \d_{ab} \bigr)(\pa \s)^6\Big) \,,
\end{align}
\begin{align}
\label{K}
K = \frac{1}{128\td{L}} \Big(& 128  d - 64 e^{2 \s} \bigl( (d-2) (\pa \s)^2 - 2  \pa^2 \s  \bigr) \nn \\
&+16 e^{4\s} \bigl( 3 (d-4)(\pa \s)^4 - 4 (\pa \s)^2 \pa^2 \s - 8  \pa_a \pa_b \s \pa^a \s \pa^b \s \bigr) \nn \\
&  -8  e^{6\s} (\pa \s)^2 \bigl( 5 (d-6)(\pa \s)^4  - 24  \pa_a \pa_b \s \pa^a \s \pa^b \s - 6 (\pa \s)^2 \pa^2 \s\bigr) \nn \\
& + 5  e^{8\s} (\pa \s)^4 \bigl( 7 (d-8)(\pa \s)^4 - 8 (\pa \s)^2 \pa^2 \s - 48  \pa_a \pa_b \s \pa^a \s \pa^b \s \bigr)\Big)\,.
\end{align}
The above results together with \eqref{Riemann}-\eqref{RicciS} satisfy the Gauss-Codazzi equation. To evaluate the total on-shell action, we need one more ingredient. Using the following expansion for a matrix $M$ in terms of its trace 
\begin{align}
\sqrt{det \,(1+M)} =  1 &+ \hlf  Tr M + \frac{1}{8}  (Tr M)^2 - \frac{1}{4}  Tr(M)^2 + \frac{1}{6}  Tr(M)^3 - \frac{1}{8} \, Tr(M^2) Tr M\nn \\
& + \frac{1}{48}  (Tr M)^3 + \frac{1}{384} \Big( (Tr M)^4\! -\! 12 (Tr M)^2 Tr(M^2) + 12 (Tr(M^2))^2\nn \\ 
& + 32\, Tr(M)^3 \,Tr M - 48\, Tr(M^4) \Big)+\cdots \,,
\end{align}
 the determinant of the boundary metric will be
\be
\sqrt{h} = \td{L}^d e^{-d \s} \bigl( 1 + \hlf  e^{ 2 \s} (\pa \s)^2 - \frac{1}{8}  e^{4\s} (\pa \s)^4 + \frac{1}{16}  e^{6\s} (\pa \s)^6 - \frac{5}{128}  e^{8\s} (\pa \s)^8+ \cdots\bigr)\,.
\ee

\section*{Acknowledgment}
A. G. would like to thanks A. Sinha for reading the manuscript and drawing our attention to ANEC subject. We would also like to thanks M. M. Sheikh-Jabbari for his valuable comments on the draft. 
This work is supported by Ferdowsi University of Mashhad under the grant 3/47298 (1397/05/16).

%%%%%%%%%%%%%%%%%%%%%%%%%%%%%%%%%%%%%%%%%%%%%%%%%%%%%%%%%%%%%
\providecommand{\href}[2]{#2}\begingroup\raggedright

\endgroup

\end{document}